\documentclass{jfm}
\usepackage[utf8]{inputenc}

\title[Bounded dissipation and velocity moments]
{Bounded dissipation law and profiles of turbulent velocity moments in wall flows}

\author[Xi Chen and Katepalli R. Sreenivasan] {Xi Chen$^{1}$ and Katepalli R. Sreenivasan$^2$}

\affiliation{$^1$ Key Laboratory of Fluid Mechanics of Ministry of Education, Beihang University (Beijing University of Aeronautics and Astronautics), Beijing, China
\\$^2$ Tandon School of Engineering, Courant Institute of Mathematical Sciences, Department of Physics, New York University, New York, USA}

\pubyear{20xx} \volume{000} \pagerange{000--000}
\date{?; revised ?; accepted ?. - To be entered by editorial office}

\usepackage{graphicx}
\usepackage{xcolor}
\usepackage{subfig}

\usepackage{authblk}
\usepackage[]{natbib}
\usepackage{numprint}
\usepackage{graphicx}
\usepackage{amssymb}
\usepackage{mathtools}
\usepackage{indentfirst}
\usepackage{multirow}

\begin{document}

\maketitle

\begin{abstract}
Turbulent wall flows offer the most direct means for understanding the effects of boundaries and viscosity on turbulent fluctuations. Available data on mean-square fluctuations in these flows show apparent contradiction with classical scaling based on the {mean} wall shear stress. We had earlier proposed an alternative model based on the principle of bounded dissipation to describe the data. Despite its putative success, a conclusive outcome requires much higher Reynolds numbers than are available at present, or can be expected to be available in the near future. However, the model can be validated satisfactorily even within the Reynolds number range already available by considering high-order moments and their distributions in the wall-normal direction. Expressions for high-order moments of streamwise velocity fluctuation $u$ are derived in the form $ \langle u^{+2q} \rangle^{1/q}=\alpha_q-\beta_q y^{\ast1/4}$; here $q$ is an integer, $\alpha_q$ and $\beta_q$ are constants independent of the friction Reynolds number $Re_\tau$, and $y^{\ast} = y/\delta$ is the distance away from the wall, normalized by the flow thickness $\delta$; in particular, $\alpha_q =\mu+\sigma q$ according to the `linear q-norm Gaussian' process, where $\mu$ and $\sigma$ are flow-independent constants. Excellent agreement is found between this formula and the available data in boundary layers, pipes and channels for $1 \leq q \leq 5$. For fixed $y^+ = y^*Re_\tau$, the present formulation leads to the bounded state $\langle u^{+2q} \rangle^{1/q}=\alpha_q$ as $Re_\tau\rightarrow\infty$. This work demonstrates the success of the present model in describing the behavior of fluctuations in wall flows.
\end{abstract}


\section{Introduction}

In the last 90 years or so, an extraordinary amount of information has become available on the dynamics of homogeneous and isotropic turbulence \cite{Frisch,SreenivasanAntonia}, as well as on passive scalars mixed by this form of turbulence \cite{Sreenivasan_mixing}. Similarly, rich information has been obtained in turbulent shear flows over a smooth wall (``wall flows") \citep{marusic2010wall}, but it is not clear how precisely the wall influences the behavior of turbulence. Questions persist \citep{Townsend1976,Smits2011ARFM} about how the mean velocity and various moments of fluctuations around the mean are distributed with distance from the wall.
The most successful paradigm that has helped organize much of the empirical information is the so-called law-of-the-wall \citep{MoninYaglom}, which posits that time averages of turbulence quantities near the wall, when suitably scaled on 
frictional stress at the wall, are invariant with respect to the Reynolds number (``Reynolds number similarity"). This scaling was advanced in the 1930s by \cite{Millikan}, who put on a firmer ground the Prandtl-Karman ``logarithmic law", according to which the mean velocity varies logarithmically with the height from the wall. Logical extensions of this principle to fluctuations yield specific forms: for example, the peak velocity fluctuation in the direction of the main motion (the so-called ``streamwise" fluctuation) is a constant independent of the Reynolds number. The convenient form of the Reynolds number here is the friction Reynolds number $Re_\tau \equiv u_\tau \delta/\nu$ where $u_\tau \equiv \sqrt{\tau_w}$, $\tau_w$ being the frictional stress at the wall in kinematic units.

As more data have accumulated in the last two decades from laboratory experiments \citep{Hultmark2012PRL,vincenti2013,LongPipe2017,Samie2018JFM,Ono2022,Ono2023} and direct numerical simulations of the Navier-Stokes equations (DNS) \citep{WuxiaohuaTBL2009,SO2010,Jimenez2010,Moser2015,Yama2018,hoyas2022,Yao2023JFM}, it has become increasingly clear that the normalized peak streamwise fluctuation intensity is an increasing function of the Reynolds number over its available range, calling into question the suitability of the law-of-the-wall for fluctuations.
An alternative possibility that has been considered is Townsend's \citep{Townsend1976} attached eddy model, whose basis is that the boundary layer dynamics is governed by eddies that remain permanently attached to the wall. The application of this hypothesis shows that
the peak variance of streamwise velocity fluctuation increases logarithmically with the Reynolds number \citep{marusic2017PRF,Marusic2019ARFM}.
On the other hand, the recent work of \cite{CS2021JFM,CS2022JFM,CS2023JFM} has shown that the increase is a finite-Reynolds-number effect, which eventually saturates to a constant, restoring the validity of the law-of-the-wall. The initial increase and eventual saturation were based on the so-called bounded dissipation model (see also \cite{Monkewitz2021}), which states that the dissipation at the wall approaches its upper bound value of 1/4 as the Reynolds number becomes very large and that all fluctuating quantities of interest are slaved to this behavior. This principle of bounded dissipation gave verifiable results for peaks of all variances of fluctuations that reside in the region close to the wall (``inner layer"), in agreement with the data.

 \begin{figure*}
 \begin{centering}
{\includegraphics[trim = 1cm 10.5cm 16.5cm 0.5cm, clip,  width=8.8 cm]{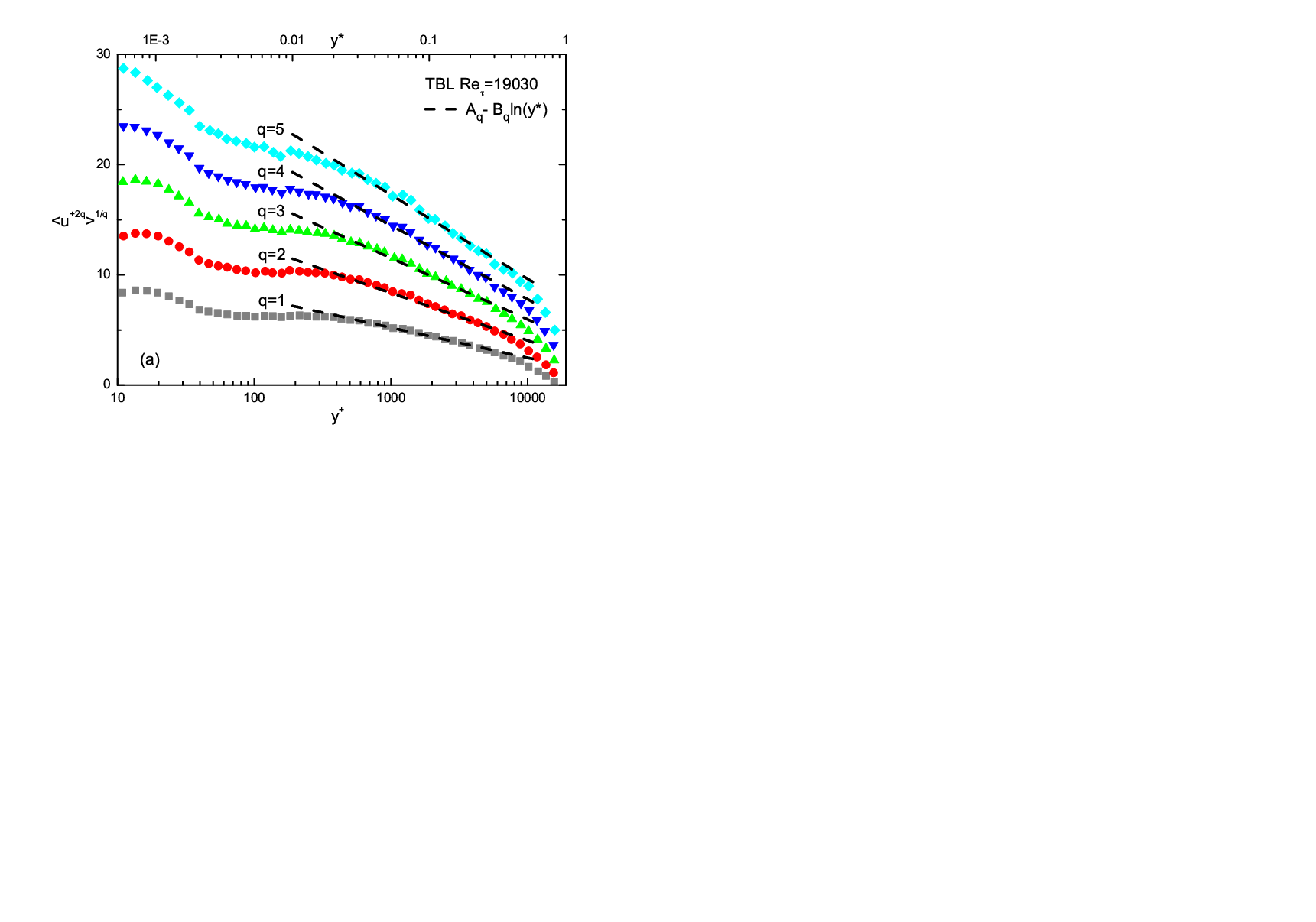}}
{\includegraphics[trim = 1cm 10.5cm 16.5cm 0.5cm, clip,  width=8.8 cm]{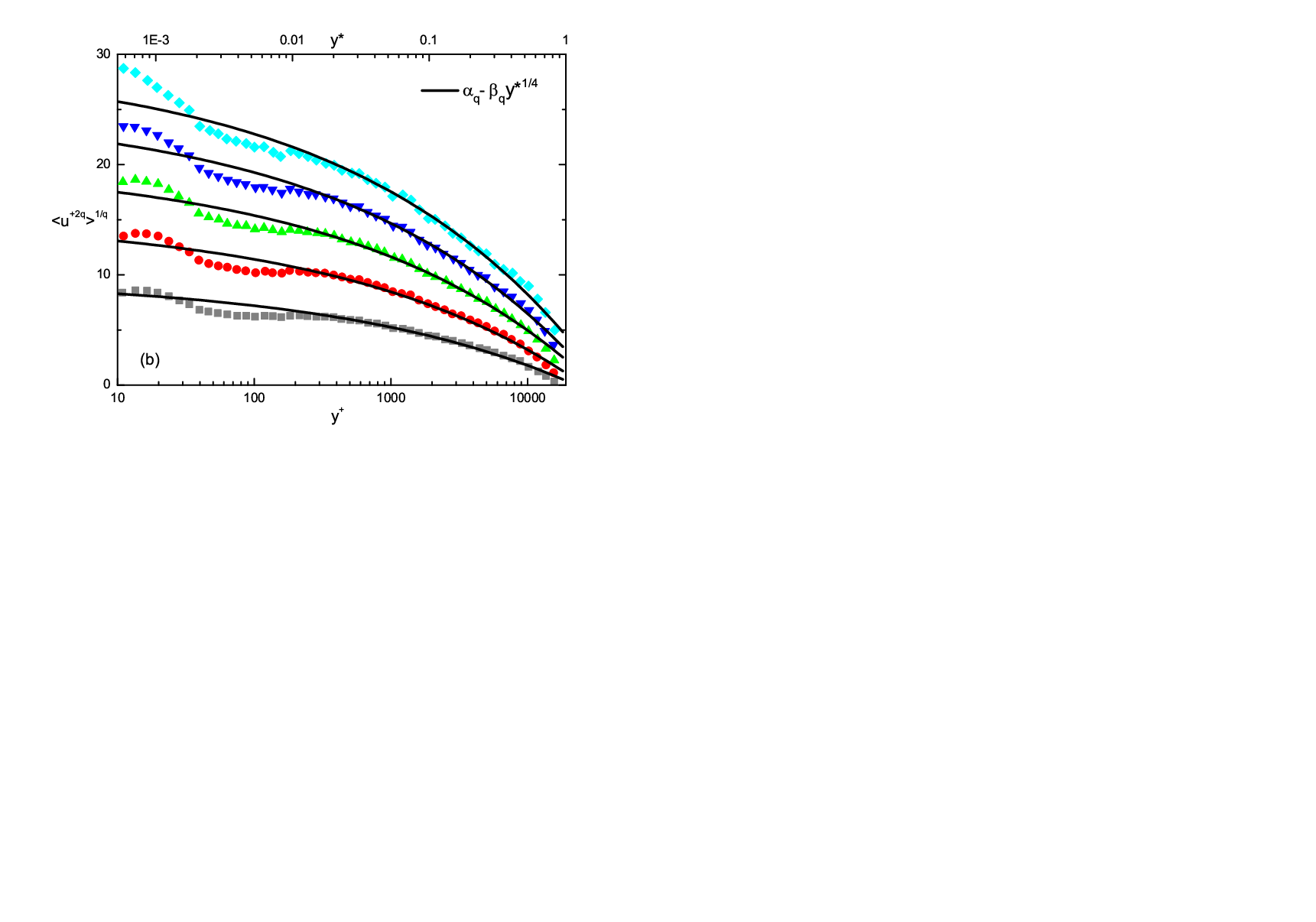}}
\caption{Wall-normal variation of $\langle u^{+2q}\rangle^{1/q}$ for $q=1-5$ in TBL compared with (a) the logarithmic variation of the attach-eddy model in Eq.~\textbf{2} by \cite{MM2013} (dashed lines), and (b) the bounded variation given by Eq.~\textbf{4} (solid lines); $q$ values are the same as in (a). In both panels, the bottom abscissa is in viscous unit $y^+=y^\ast/Re_\tau$, and the top in the outer unit $y^\ast=y^+/Re_\tau$. Symbols are data measured by \cite{Hutchins2009}.}
\label{fig:u2p:TBL}
\end{centering}
\end{figure*}

The underpinnings of both models are not foolproof at present (see below for more comments), so agreement with empirical results is a more solid basis for deciding the relative merits of the models. It has been pointed out, rightly, that the available Reynolds number range is not adequate to make definitive claims in favor of one model or the other: comparison of only velocity variances in the available data range is ambiguous in some instances---though independent assessments \citep{nagib2024utilizing,Pirozzoli2024}, besides our own \citep{CS2021JFM,CS2022JFM,CS2023JFM}, document support for the bounded law over the attached eddy results.
Thus, a detailed consideration of high-order velocity moments is essential. Here, we consider not only high-order moments but also the wall-normal profiles of such moments up to order 10, and make detailed data comparisons for the two formulations. We show that the bounded dissipation model is better supported by the data, and follow it up with some general conclusions. This defines the scope of this article.

\section{Brief theoretical considerations}

{\textit {A.~Attached eddy model:}} According to attached eddy model, the variance of streamwise velocity $u$ follows a logarithmic decay as
\begin{equation}\label{eq:townsend}
\langle u^{+2} \rangle = B_1 -A_1 \ln(y^\ast).
\end{equation}
Here, the superscript $+$ indicates normalization by $u_\tau$ and $y^\ast=y/\delta$.
The brackets $\langle \rangle$ indicate temporal averages. This equation is meant to be valid in the overlap region, nominally associated with the logarithmic distribution of the mean velocity profile \citep{Hultmark2012PRL,MM2013}. The same reference \cite{MM2013} generalizes this formula for high-order moments 
as
\begin{equation}
\label{eq:M&M}
\langle u^{+2q} \rangle^{1/q} = B_q -A_q \ln(y^\ast).
\end{equation}
Here $A_q$ and $B_q$ are constants independent of $y^\ast$ and $Re_\tau$ but depend on the moment order $q$. In going from Eq.~\textbf{1} to Eq.~\textbf{2}, a Gaussian distribution of $u$ has been assumed, leading to $\langle u^{+2q}\rangle^{1/q} =[(2q-1)!!]^{1/q}\langle u^{+2} \rangle$ with $q!!\equiv q(q-2)(q-4)\ldots1$, though it is known from \cite{MM2013} that the slopes $A_q$ deviate significantly from the Gaussian expectation. Analogous to the Karman constant\footnote{We are aware that the Karman constant may depend on the flow \citep{nagib2007,Monkewitz_Nagib_2023}.} in the mean velocity distribution, quantities $A_q$ are called the Townsend-Perry constants \citep{Marusic2019ARFM,Morrison2019,MM2013,Bjorn2021PRR,Smits2022prize}.

Equation~\textbf{1} builds on the notion that turbulent eddies are effectively attached to the wall, with their number density varying inversely with $y^\ast$ \citep{Marusic2019ARFM}. Because the identification of a clear eddy structure is fraught with uncertainties \citep{Schlatter2014}, the {$k^{-1}$ velocity spectrum \citep{Perry1986}} is sometimes thought to be another possible rationale for the same idea; but that, too, has not been observed unambiguously \citep{vallikivi2015,Panton2017}. To avoid these problems, \cite{Hultmark2012JFM} simply invoked an overlap argument. Thus, there is considerable {\it a priori} room for the debate on the logarithmic profile for the fluctuation intensity.

{\textit {B. Bounded dissipation model:}} According to \cite{CS2021JFM,CS2022JFM,CS2023JFM}, the alternative model based on the boundedness of dissipation yields
\begin{equation}\label{eq:C&S1}
    \langle u^{+2} \rangle = \alpha_1 -\beta_1 (y^\ast)^{1/4}.
\end{equation}
The procedure is summarized here only briefly because the details can be found in \cite{CS2023JFM}. We have shown that an inner expansion of the form $\langle u^{+2} \rangle=f_0(y^+)+f_1(y^+) g(Re_\tau)$ works very well close to the peak fluctuations. Here, $y^+=y^\ast Re_\tau$ is the inner viscous scale, and the Gauge function $g(Re_\tau)=Re^{-1/4}_\tau$ depicts the finite $Re_\tau$ dependence. Similarly, the appropriate  outer expansion takes the form $\langle u^{+2} \rangle=F_0(y^\ast)$.
Matching the two expansions results in Eq.~\textbf{3}.  Similar matching of inner-outer expansions of a high-order moment $q$ leads to
\begin{equation}
\label{eq:C&S}
\langle u^{+2q} \rangle^{1/q} = \alpha_q -\beta_q (y^\ast)^{1/4},
\end{equation}
where $\alpha_q$ and $\beta_q$ are constants independent of $y^\ast$ and $Re_\tau$.

In contrast to Millikan's matching for the logarithmic mean velocity, the procedure here matches the two-term inner expansion with the full outer scaling. Equation~\textbf{4} could then be considered a generalization of Eq.~\textbf{3} by replacing $\langle u^{+2} \rangle$ with $\langle u^{+2q} \rangle^{1/q}$. Another way to derive Eq.~\textbf{4} is to invoke a ``linear-q-norm Gaussian'' (LQNG) process developed in \cite{CS2022JFM}, which is different from the derivation of Eq.~\textbf{2} on the basis of the Gaussian assumption. This will be explained more later.

\begin{figure*}
\begin{centering}
{\includegraphics[trim = 1cm 11.2cm 16.cm 1cm, clip,  width=6.5 cm]{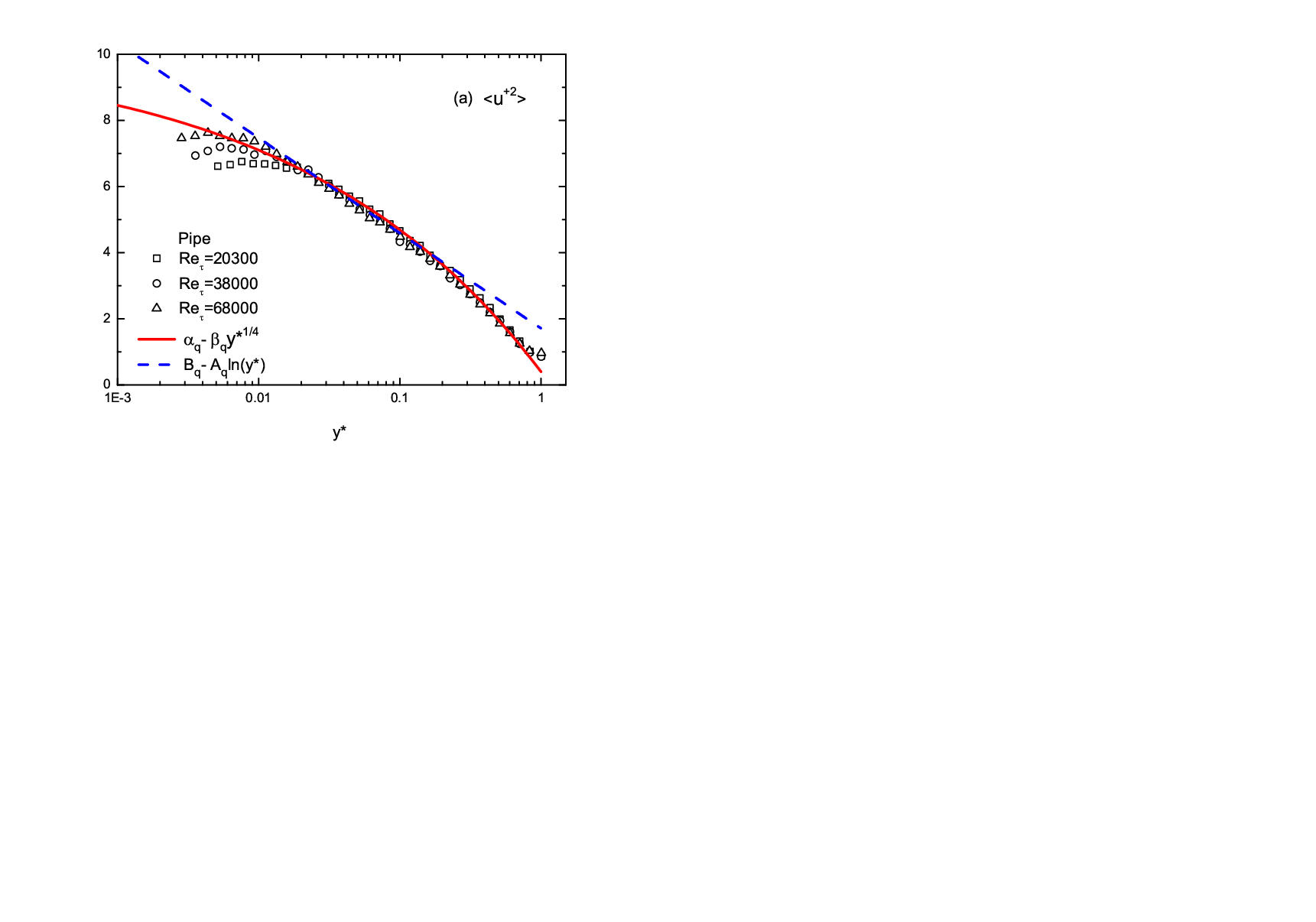}}
{\includegraphics[trim = 1cm 11.2cm 16.cm 1cm, clip,  width=6.5 cm]{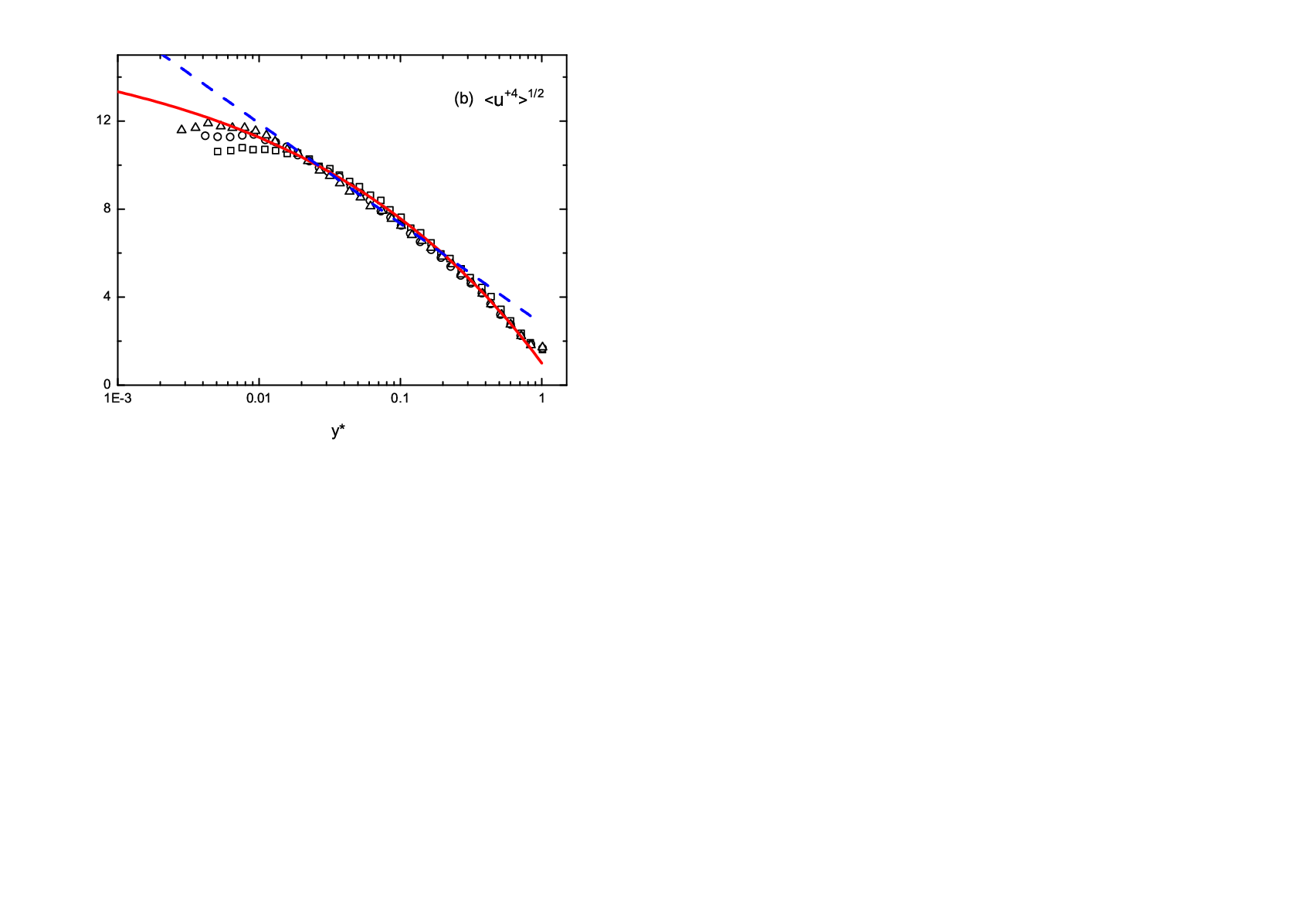}}\\
{\includegraphics[trim = 1cm 10.5cm 16.cm 1cm, clip,  width=6.5 cm]{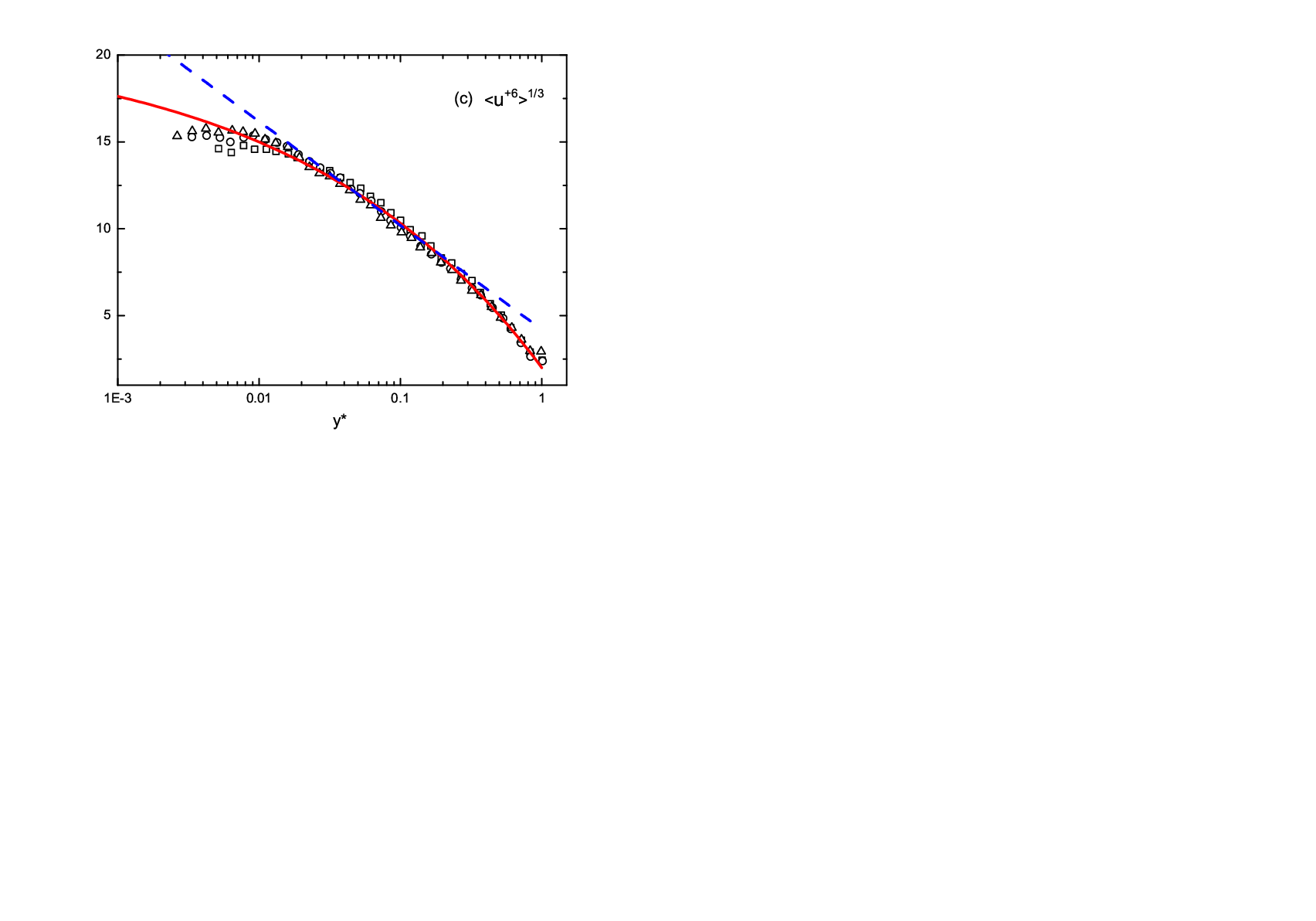}}
{\includegraphics[trim = 1cm 10.5cm 16.cm 1cm, clip,  width=6.5 cm]{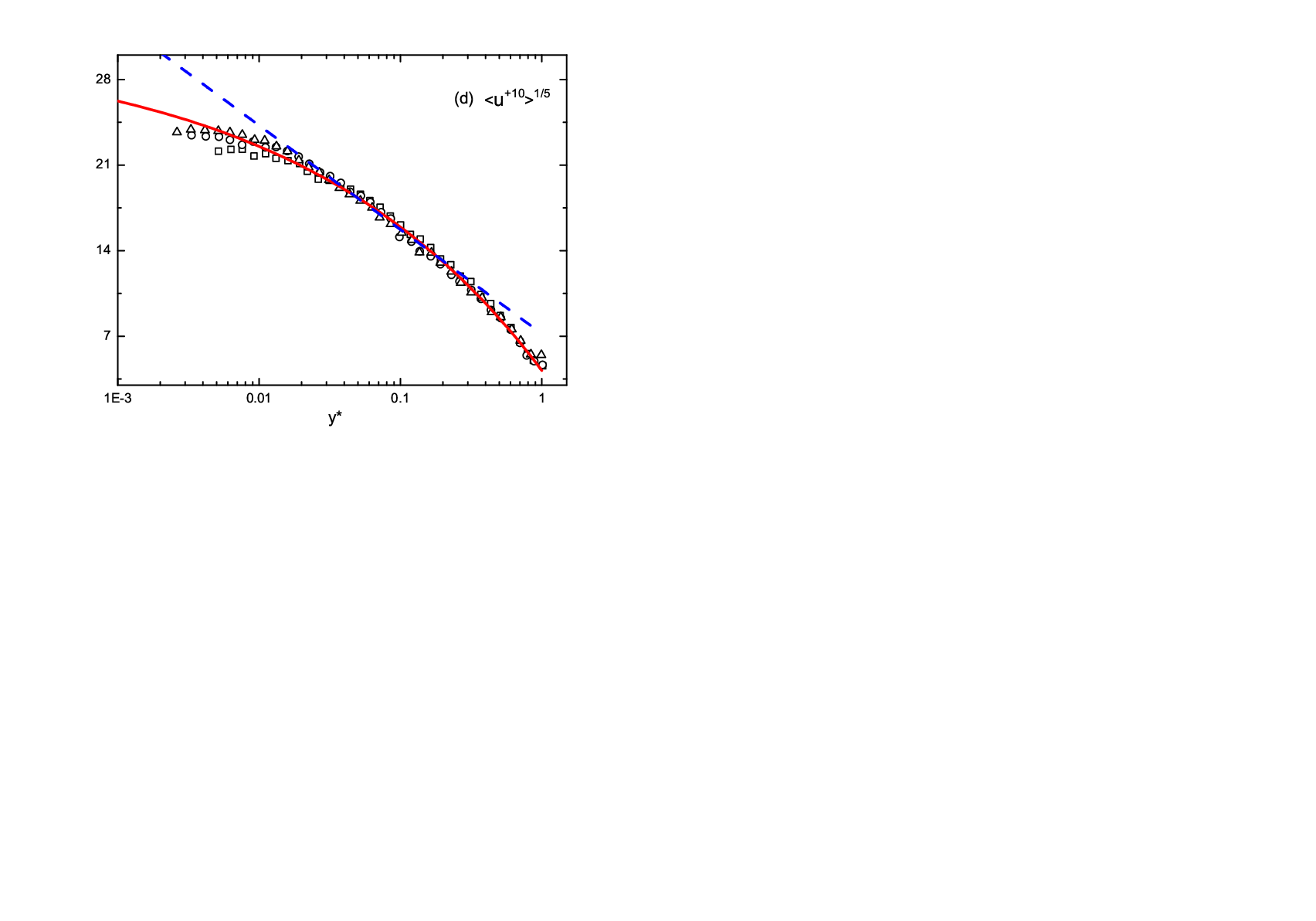}}\\
\caption{Wall-normal variations of $\langle u^{+2p}\rangle^{1/p}$ in pipes for (a) $p=1$; (b) $p=2$; (c) $p=3$ and (d) $p=5$. Solid lines (red) are the bounded variation in Eq.~\textbf{4}, whilst dashed lines (blue) are the logarithmic variation in Eq.~\textbf{2}. Symbols are experimental data from \cite{Hultmark2013}.}
\label{fig:u2p:Pipe}
\end{centering}
\end{figure*}

\section{Comparison with experiments}
Below, side-by-side performances of Eq.~\textbf{2} and Eq.~\textbf{4} will be compared with the data and variations of $\alpha_q$ and $A_q$ will be examined to understand their underlying statistical properties in the asymptotic limit of $Re_\tau \to \infty$.

Reminiscent of the standard overlap in the form of the well-known log-law, the generalized form of Eq.~\textbf{2} was proposed in \cite{MM2013} for an intermediate region which, for high end of the available Reynolds numbers, is approximately in the range $400 < y^+<0.3Re_\tau$. For a decade of $y^+$ in this intermediate region, an $Re_\tau>13,000$ is required (see also \cite{nagib2024utilizing}), which is larger than those available from DNS. We therefore start with comparisons for high-$Re_\tau$ experiments of TBL and pipe flows, and subsequently consider the DNS data for channels obtained at a more moderate $Re_\tau$. Uncertainties in the data are not addressed here, because they have been discussed in the original references.

Specifically, Fig.~\ref{fig:u2p:TBL} shows $\langle u^{+2q}\rangle^{1/q}$ of TBL at $Re_\tau = 19,000$ for $q=1-5$, measured by \cite{Hutchins2009} in the Melbourne wind tunnel. This same data set is used in \cite{MM2013} to illustrate the generalized representation of Eq.~\textbf{2}. Dashed lines in Fig.~\ref{fig:u2p:TBL}a are the best fits given by Eq.~\textbf{2}; there is decent agreement with the data in the intermediate $y^+$ range, with the best fits for $(A_q, B_q)$ given by: $(1.19, 1.71)$, $(2.06, 2.50)$, $(2.93, 3.23)$, $(3.81, 4.12)$ and $(4.68, 5.03)$ for $q=1-5$, respectively. {Both the slopes $A_p$ and the intercepts $B_q$} increase with increasing $q$, but a closer look at Fig.~\ref{fig:u2p:TBL}a reveals that the inner limits for the fits of Eq.~\textbf{2} commence increasingly farther away from the wall---for example, $y^+ \approx 200$ for $q=1$ compared to $y^+ \approx 400$ for $q=5$. The expected logarithmic region will erode continually if this trend persists. We should also note that the fit becomes more of a tangent to the data for larger $q$.

\begin{figure*}
\begin{centering}
{\includegraphics[trim = 1.5cm 10.5cm 16.cm 1cm, clip,  width=8.8 cm]{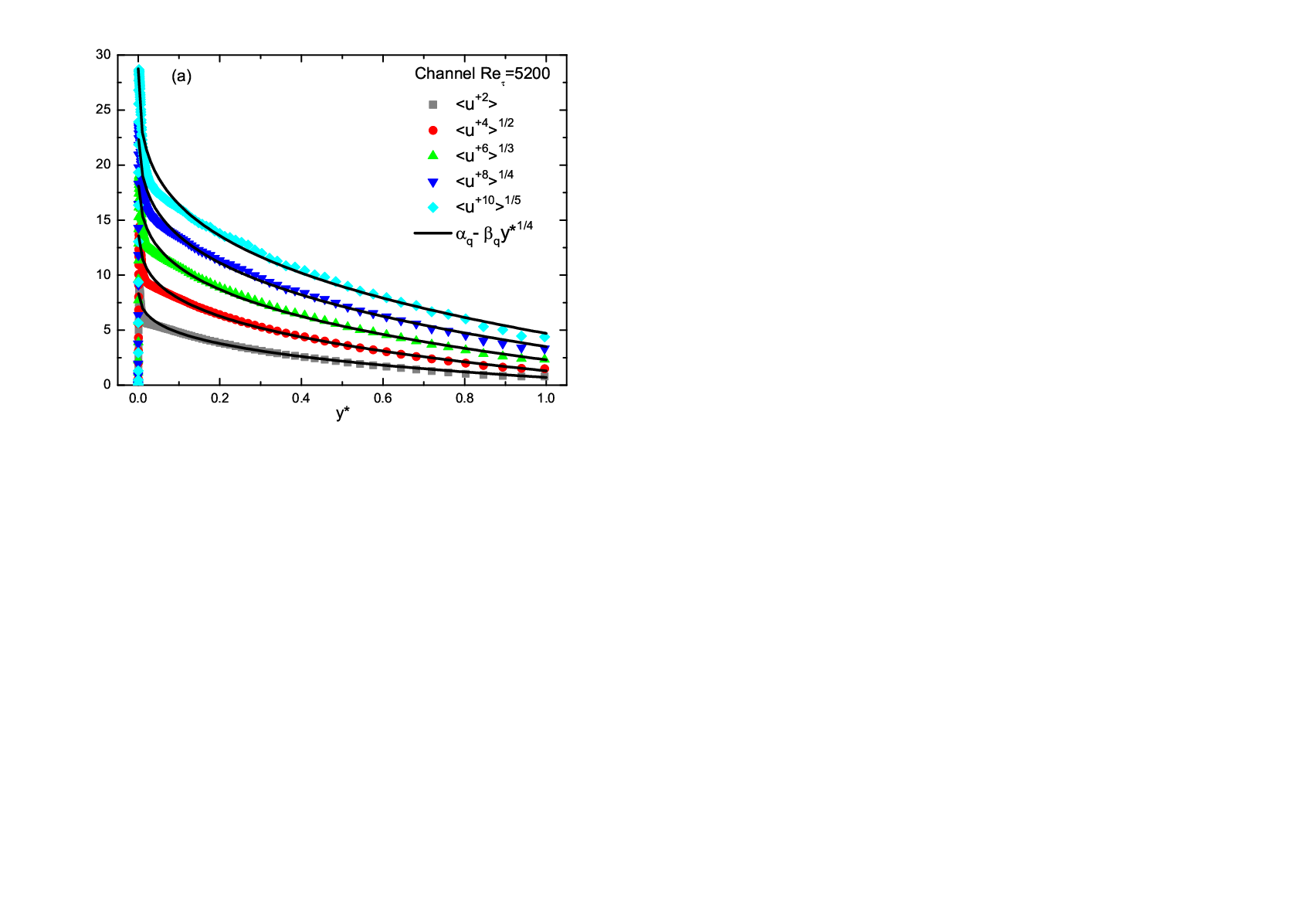}}
{\includegraphics[trim = 1.5cm 10.5cm 16.cm 1cm, clip,  width=8.8 cm]{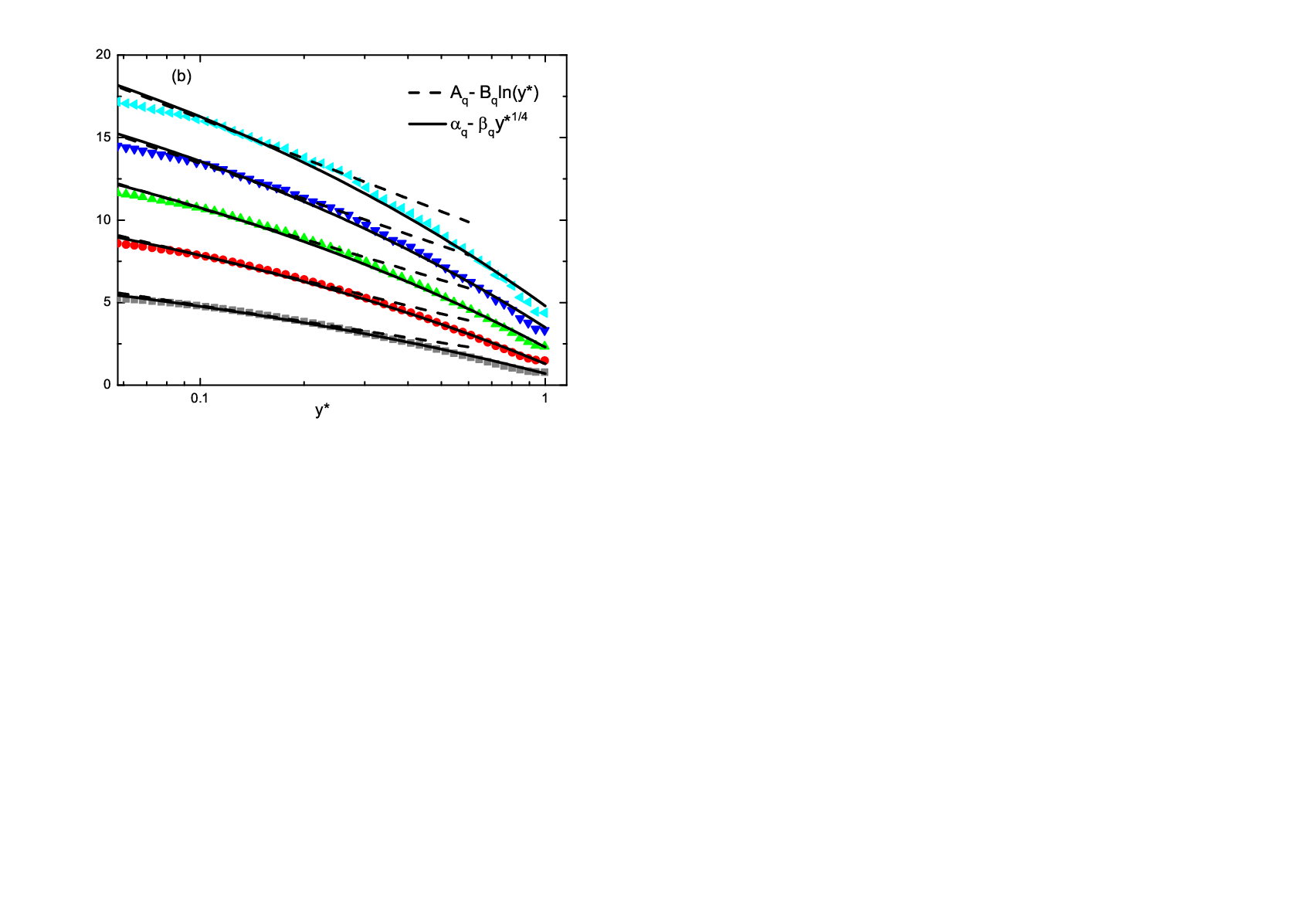}}
\caption{Wall-normal variation of {$\langle u^{+2q}\rangle^{1/q}$} in the channel for {$q=1-5$}: abscissa in linear scale (a) and in logarithmic scale (b). Symbols are DNS data from {\cite{Moser2015}}. Solid lines are Eq.~\textbf{4} while dashed lines are Eq.~\textbf{2}.}
\label{fig:u2p:CH}
\end{centering}
\end{figure*}

On the other hand, the bounded behavior of Eq.~\textbf{4} reproduces the data better over a wider flow domain (Fig.~\ref{fig:u2p:TBL}b), i.e., from $y^+=200$ to almost the edge of the boundary layer. The best fits for $(\alpha_q, \beta_q)$ are $(9.7, 9.3)$, $(15.2, 14.1)$, $(20.2, 17.9)$, $(25.2, 22)$ and $(29.5, 25)$ for $q=1-5$, respectively. We can make two other comments. First, since $\langle u^{+2q} \rangle^{1/q}$ remains positive at the edge of the flow (i.e., for $y^\ast=1$), $\beta_q$ is by necessity slightly smaller than $\alpha_q$. A corollary is that the fluctuation moments at the edge of the flow are constants independent of $Re_\tau$. Second, $\langle u^{+2q} \rangle^{1/q}=\alpha_q$ is the asymptotic plateau of $y^\ast\rightarrow0$, which can therefore be used to approximate the near-wall data. For example, closer to the wall within the so-called buffer layer (i.e. $y^+\approx40$), Eq.~\textbf{4} approximates the data moderately well. In fact, the deviation from the data is $O(1)$ for the entire flow domain. However, this is not the case for the attached eddy model \citep{marusic2017PRF} for which the peak scales as $(A_q/2) \ln Re_\tau$, while the near-wall extension of Eq.~\textbf{2} suggests a magnitude $A_q \ln Re_\tau$; the difference between the two is of order $\ln Re_\tau$ which diverges as $Re_\tau\rightarrow\infty$. Hence, one cannot simply extrapolate Eq.~\textbf{2} to approximate the near-wall profile.

Moving on to the pipe flow, Figs.~\ref{fig:u2p:Pipe}a-d compare the profiles of $\langle u^{+2p}\rangle^{1/p}$ for $q=1,2,3,5$ with data measured by \cite{Hultmark2013} in the Princeton Superpipe (see also \cite{vallikivi2015}). Taking data for $Re_\tau=20,300$ as an example, the logarithmic fit is obtained in the range $0.02<y^\ast<0.3$, which corresponds to $400<y^+<0.3Re_\tau$. For that range, we obtain $(A_q, B_q)$ as $(1.25, 1.71)$, $(1.98, 2.78)$, $(2.6, 4.20)$ and $(3.7, 7.22)$ for $q=1,2,3,5$, respectively. Again, they appear somewhat like tangents to the data profiles. In comparison, the bounded form agrees closely with data in the whole range $y^\ast>0.02$, with the best fits of $(\alpha_q, \beta_q)$ given by  $(10.2, 9.8)$, $(16, 15)$, $(21, 19)$, $(31, 26.8)$ for increasing $q$. Note that the same constants produce good data collapse for all $Re_\tau$ profiles. This collapse is the basis for our outer expansion mentioned in the section on ``Brief theoretical considerations". It is clear in Fig.~2 that the form based on bounded dissipation extends almost to the center of the pipe, much farther than the logarithmic form that does not extend beyond $y^\ast=0.3$. Also, towards the wall, the bounded form characterizes larger segments of the flow.

Comparisons with channel data are shown in Fig.~3 for the DNS data at $Re_\tau=5200$ by \cite{Moser2015}. These high-order moments are obtained from their data maintained at the Johns Hopkins Turbulence Database. More than $5 \times 10^6$ velocity samples are averaged to obtain these profiles, and the convergence of the probability density function has been verified. As shown in Fig.~3a, the bounded form yields a close representation of the data, covering both the near-wall and the center flow regions. For a closer look near the wall, Fig.~3b shows the same plot on the logarithmic abscissa, from which it is clear that the logarithmic form only captures data in a narrow domain (from $y^+=400$ or $y^\ast=0.08$, to $y^\ast=0.3$), while Eq.~\textbf{4} offers a better description.

\section{Asymptotic behaviors}
Different asymptotic behaviors can be deduced for the two proposals. Unlike the unbounded logarithmic growth of variance according to Eq.~\textbf{1}, a bounded state follows for Eq.~\textbf{3}, i.e., $\langle u^{+2} \rangle=\alpha_1\approx10$ as $Re_\tau\rightarrow\infty$. This feature also appears for higher $q$. It is therefore important to understand how the $A_q$ and $\alpha_q$ vary with $q$ because they form the leading terms in the two models. They are plotted in Fig.~4 for all three flows.

\begin{figure*}
\begin{centering}
{\includegraphics[trim = 1.5cm 10.5cm 16.cm 1cm, clip,  width=8.8 cm]{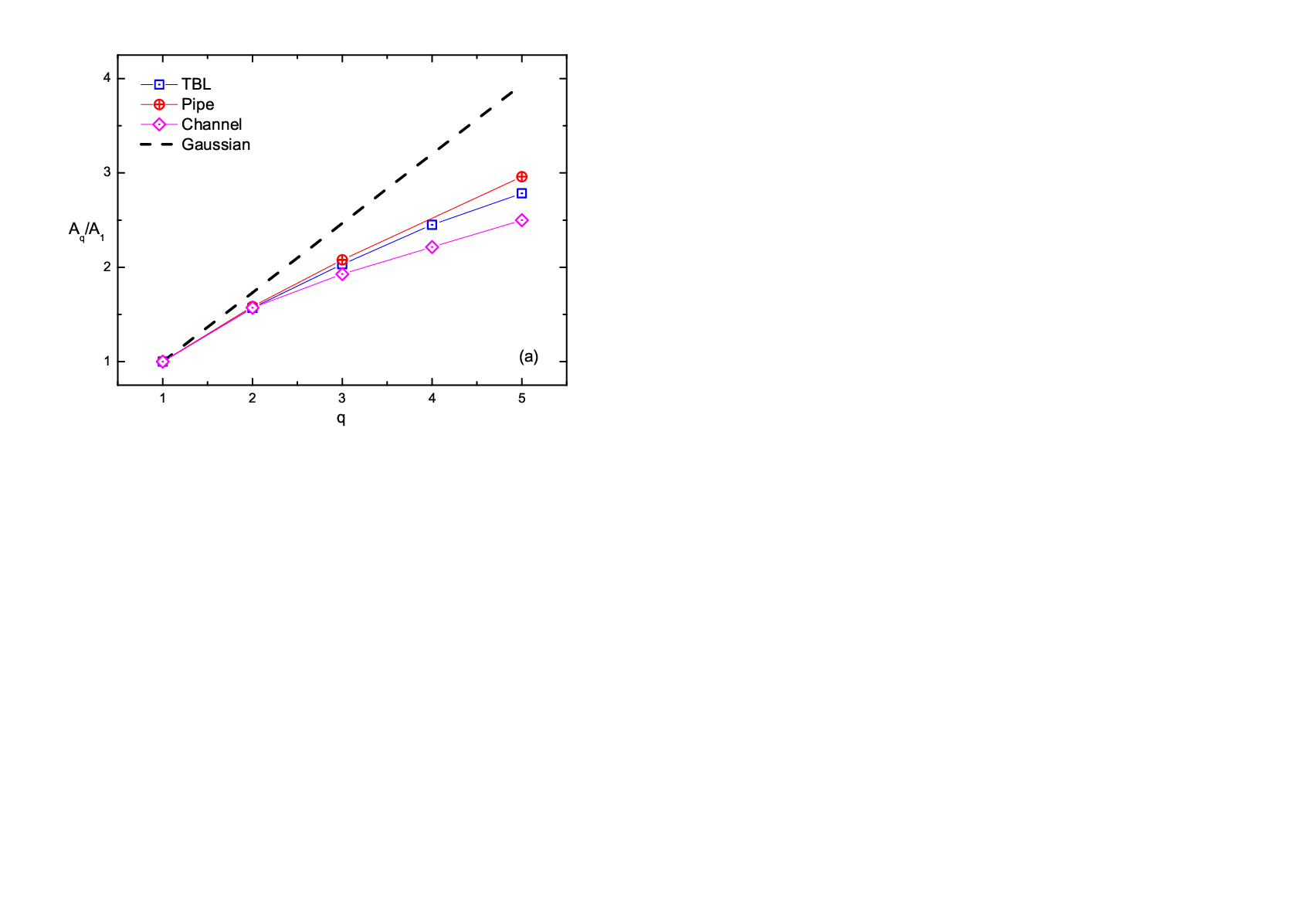}}
{\includegraphics[trim = 1.5cm 10.5cm 16.cm 1cm, clip,  width=8.8 cm]{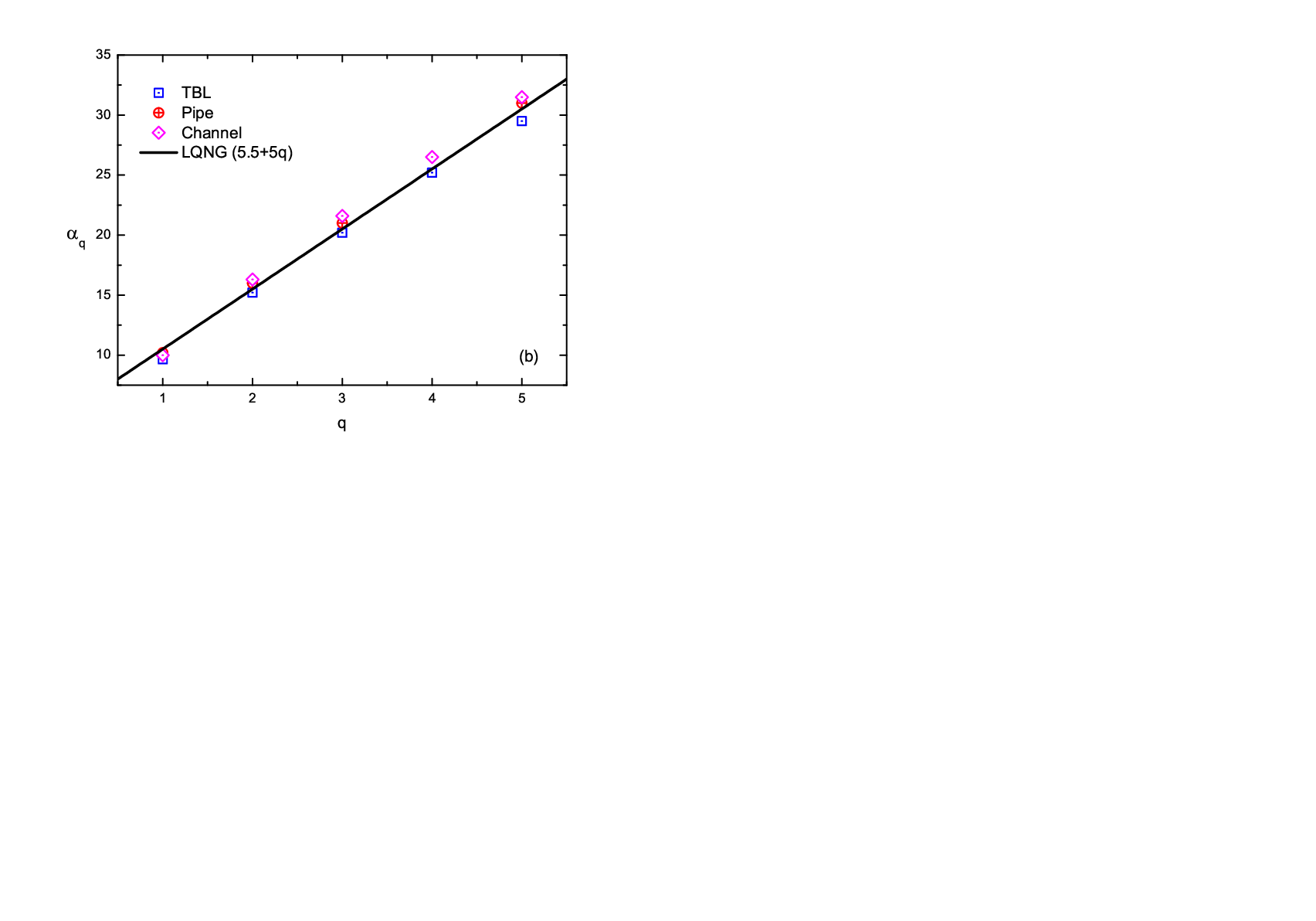}}
\caption{(a) Ratio of $A_q/A_1$ varying with order $q$; dashed line represents the Gaussian prediction by Eq.~\textbf{5}. (b) $\alpha_q$ varying with order $q$; solid line indicates the LQNG process in Eq.~\textbf{7}.}
\label{fig:u2p:CH}
\end{centering}
\end{figure*}

For $A_q$ (normalized by $A_1$), the data vary with the flow, those for the channel being the lowest. This is in addition to the sensitivity shown by $A_1$ itself with respect to $Re_\tau$ \citep{Morrison2019}. Moreover, following the Gaussian model in \cite{MM2013}, one has
\begin{equation}\label{eq:ApGaussain}
A_q=A_1[(2q-1)!!]^{1/q},
\end{equation}
which is, however, conspicuously higher than the $A_q$ obtained from Fig.~4a. This departure indicates that the Gaussian assumption does not hold and that a different model is needed (e.g., the sub-Gaussian consideration of \cite{Bjorn2021PRR}).

In contrast, the $\alpha_q$ values for the three flows are quite close to each other and exhibit an excellent linear dependence with the order $q$ (Fig.~4b). As $\langle u^{+2q} \rangle^{1/q}=\alpha_q$ for $y^\ast\rightarrow0$, the linear growth of $\alpha_q$ actually corresponds to a LQNG process of $\phi=(u^+)^2$, which satisfies the operator reflection symmetry \citep{CS2022JFM} as
\begin{equation}\label{eq:LPNG}
\mathbf{E}\circ\mathbf{Q}(\phi)=\mathbf{Q}\circ\mathbf{E}(\kappa).
\end{equation}
Here, $\kappa=N(\mu, {2\sigma})$ is a Gaussian seed with mean $\mu$ and variance $2\sigma$; $\mathbf{E}$ is the exponential transform operator, e.g. $\mathbf{E}(\kappa)=e^{\kappa}$; and $\mathbf{Q}$ is the operator of $q$-norm estimation, e.g. $\mathbf{Q}(\phi)=\langle\phi^{q} \rangle^{1/q}$ with $\langle \cdot \rangle$ for the expectation. To see the linearity with $q$, applying $\mathbf{E}^{-1}$ (i.e., taking the logarithm) on both sides of Eq.~\textbf{6} yields
\begin{eqnarray}\label{eq:LPNG:u}
\langle u^{+2q}\rangle^{1/q} = \mathbf{Q}(\phi)&=&\mathbf{E}^{-1}\circ\mathbf{Q}\circ\mathbf{E}(\kappa)\nonumber\\
& = & \ln[\langle e^{\kappa q} \rangle^{1/q}]
=\mu  +\sigma q.
\end{eqnarray}
With $\mu=5.5$ and $\sigma=5$, Eq.~\textbf{7} explains the linear behavior of $\alpha_q$ in Fig.~4b. Note that the present relation $\alpha_q = 5.5 + 5q$ is very close to $\alpha_q = 5.5 + 5.9q$ from Eq.~(4.10) of \cite{CS2022JFM} for the inner peak moments at $y^+=15$, indicating that the entire flow shares the same LQNG property.

The LQNG property can now be used to derive Eq.~\textbf{4} from Eq.~\textbf{3}. According to Eq.~\textbf{6} or Eq.~\textbf{7}, the ratio of moments increases linearly with order, i.e., $\langle u^{+2q}\rangle^{1/q}/\langle u^{+2}\rangle=(\mu+\sigma q)/(\mu+\sigma)=c_1+c_2 q$ where $c_1=\mu/(\mu+\sigma)$ and $c_2=\sigma/(\mu+\sigma)$. Furthermore, with Eq.~\textbf{3} for $\langle u^{+2}\rangle$, one obtains Eq.~\textbf{4} for high-order moments. It is evident that LQNG describes Fig.~4b quite consistently.

\section{Conclusions}
The most important quantity in turbulent wall flows is the mean velocity distribution and the associated gradient at the wall. For these flows with large fluctuations, it is natural that they will affect the mean flow strongly. Thus, there is considerable interest in the behavior of fluctuations in wall flows. For some time now, the attached eddy model has been advanced \citep{Marusic2019ARFM} as a suitable formulation to explain observed behaviors (experiment and DNS). Recent work based on bounded dissipation \citep{CS2021JFM,CS2022JFM,CS2023JFM} offers a better alternative as far as it can be judged by comparison with the data. In this paper, we subject the latter alternative to more rigorous tests by comparing high-order moments (instead of restricting simply to the variance) and also the wall-normal profiles of these high-order moments. We find that the bounded dissipation paradigm for high-order moments of wall turbulence is in excellent agreement with the data of boundary layers, pipes and channels. The new expression covers a wide wall-normal range and suggests a constant plateau for the moments that grow linearly with the order of the moment. Compared to the attached eddy model, the current results support the recovery of the classical law-of-the-wall for asymptotically high Reynolds number.

Future work will focus on the connection of the present formulation to the flow geometry.

\emph{Acknowledgement.} 
{We are grateful to all the authors cited in figures 1-3 for making their data available. X. Chen appreciates the support by the National Natural Science Foundation of China, No. 12072012 and 11721202, and the ``Fundamental Research Funds for the Central Universities".}

\bibliography{pnas-sample}

\begin{thebibliography}{42}
\expandafter\ifx\csname natexlab\endcsname\relax\def\natexlab#1{#1}\fi

\bibitem[Birnir {\em et~al.\/}(2021)Birnir, Angheluta, Kaminsky \&
  Chen]{Bjorn2021PRR}
{\sc Birnir, Bj\"orn, Angheluta, Luiza, Kaminsky, John \& Chen, Xi} 2021
  Spectral link of the generalized {T}ownsend-{P}erry constants in turbulent
  boundary layers. {\em Phys. Rev. Res.\/} {\bf 3}, 043054.

\bibitem[Chen \& Sreenivasan(2021)]{CS2021JFM}
{\sc Chen, X. \& Sreenivasan, K.~R.} 2021 Reynolds number scaling of the peak
  turbulence intensity in wall flows. {\em J. Fluid Mech.\/} {\bf 908}, R3.

\bibitem[Chen \& Sreenivasan(2022)]{CS2022JFM}
{\sc Chen, X. \& Sreenivasan, K.~R.} 2022 Law of bounded dissipation and its
  consequenes in turbulent wall flows. {\em J. Fluid Mech.\/} {\bf 933}, A20.

\bibitem[Chen \& Sreenivasan(2023)]{CS2023JFM}
{\sc Chen, X. \& Sreenivasan, K.~R.} 2023 Reynolds number asymptotics of
  wall-turbulence fluctuations. {\em J. Fluid Mech.\/} {\bf 976}, A21.

\bibitem[Diwan \& Morrison(2019)]{Morrison2019}
{\sc Diwan, S.S. \& Morrison, J.~F.} 2019 Reynolds-number dependence of the
  {T}ownsend-perry `constant' in wall turbulence. {\em 11th International
  Symposium on Turbulence and Shear Flow Phenomena (TSFP11)\/} {\bf 728},
  Southampton, UK.

\bibitem[Frisch(1995)]{Frisch}
{\sc Frisch, Uriel} 1995 {\em Turbulence: the legacy of AN Kolmogorov\/}.
  Cambridge university press.

\bibitem[Hoyas {\em et~al.\/}(2022)Hoyas, Oberlack, Alcantara-Avila,
  Kraheberger \& Laux]{hoyas2022}
{\sc Hoyas, S., Oberlack, M., Alcantara-Avila, F., Kraheberger, S.V. \& Laux,
  J.} 2022 Wall turbulence at high friction {R}eynolds numbers. {\em Phys. Rev.
  Fluids\/} {\bf 7}, 014602.

\bibitem[Hultmark(2012)]{Hultmark2012JFM}
{\sc Hultmark, M.} 2012 A theory for the streamwise turbulent fluctuations in
  high {R}eynolds number pipe flow. {\em J. Fluid Mech.\/} {\bf 707}, 575--584.

\bibitem[Hultmark {\em et~al.\/}(2012)Hultmark, Vallikivi, Bailey \&
  Smits]{Hultmark2012PRL}
{\sc Hultmark, M., Vallikivi, M., Bailey, S.C.C. \& Smits, A.J.} 2012 Turbulent
  pipe flow at extreme {R}eynolds numbers. {\em Phys. Rev. Lett.\/} {\bf 108},
  094501.

\bibitem[Hultmark {\em et~al.\/}(2013)Hultmark, Vallikivi, Bailey \&
  Smits]{Hultmark2013}
{\sc Hultmark, M., Vallikivi, M., Bailey, S. C.~C. \& Smits, A.} 2013
  Logarithmic scaling of turbulence in smooth- and rough-wall pipe flow. {\em
  J. Fluid Mech.\/} {\bf 728}, 376–395.

\bibitem[Hutchins {\em et~al.\/}(2009)Hutchins, Nickels, Marusic \&
  Chong]{Hutchins2009}
{\sc Hutchins, N., Nickels, T.B., Marusic, I. \& Chong, M.S.} 2009 Hot-wire
  spatial resolution issues in wall-bounded turbulence. {\em J. Fluid Mech.\/}
  {\bf 635}, 103--136.

\bibitem[Jimenez {\em et~al.\/}(2010)Jimenez, Hoyas, Simens \&
  Mizuno]{Jimenez2010}
{\sc Jimenez, J., Hoyas, S., Simens, M.P. \& Mizuno, Y.} 2010 Turbulent
  boundary layers and channels at moderate {R}eynolds numbers. {\em J. Fluid
  Mech.\/} {\bf 657}, 335--360.

\bibitem[Lee \& Moser(2015)]{Moser2015}
{\sc Lee, M. \& Moser, R.D.} 2015 Direct numerical simulation of turbulent
  channel flow up to {R}e$_\tau$ = 5200. {\em J. Fluid Mech.\/} {\bf 774},
  395--415.

\bibitem[Marusic {\em et~al.\/}(2017)Marusic, Baars \&
  Hutchins]{marusic2017PRF}
{\sc Marusic, I., Baars, W.J. \& Hutchins, N.} 2017 Scaling of the streamwise
  turbulence intensity in the context of inner-outer interactions in wall
  turbulence. {\em Phys. Rev. Fluids\/} {\bf 2}, 100502.

\bibitem[Marusic {\em et~al.\/}(2010)Marusic, McKeon, Monkewitz, Nagib, Smits
  \& Sreenivasan]{marusic2010wall}
{\sc Marusic, I., McKeon, B.J., Monkewitz, P.A., Nagib, H.M., Smits, A. \&
  Sreenivasan, K.~R.} 2010 Wall-bounded turbulent flows at high {R}eynolds
  numbers: {R}ecent advances and key issues. {\em Phy. Fluids\/} {\bf 22},
  065103.

\bibitem[Marusic \& Monty(2019)]{Marusic2019ARFM}
{\sc Marusic, I. \& Monty, J.P.} 2019 Attached eddy model of wall turbulence.
  {\em Annu. Rev. Fluid Mech.\/} {\bf 51}, 49--74.

\bibitem[Meneveau \& Marusic(2013)]{MM2013}
{\sc Meneveau, C. \& Marusic, I.} 2013 Generalized logarithmic law for
  high-order moments in turbulent boundary layers. {\em J. Fluid Mech.\/} {\bf
  719}, R1.

\bibitem[Millikan(1938)]{Millikan}
{\sc Millikan, C.M.} 1938 A critical discussion of turbulent flows in channels
  and circular tubes. {\em In Proc. 5th Intl Congress of Applied Mechanics.
  John Wiley and Sons.\/} pp. 386--392.

\bibitem[Monin \& Yaglom(1971)]{MoninYaglom}
{\sc Monin, A.S. \& Yaglom, A.M.} 1971 {\em Statistical Fluid Mechanics, Volume
  I: Mechanics of Turbulence\/}. MIT Press, Cambridge, Massachusetts.

\bibitem[Monkewitz(2022)]{Monkewitz2021}
{\sc Monkewitz, P.A.} 2022 Asymptotics of streamwise {R}eynolds stress in wall
  turbulence. {\em J. Fluid Mech.\/} {\bf 931}, A18.

\bibitem[Monkewitz \& Nagib(2023)]{Monkewitz_Nagib_2023}
{\sc Monkewitz, P.A. \& Nagib, H.M.} 2023 The hunt for the {K}ármán
  ‘constant’ revisited. {\em Journal of Fluid Mechanics\/} {\bf 967}, A15.

\bibitem[Nagib {\em et~al.\/}(2007)Nagib, Chauhan \& Monkewitz]{nagib2007}
{\sc Nagib, H.M., Chauhan, K.A. \& Monkewitz, P.A.} 2007 Approach to an
  asymptotic state for zero pressure gradient turbulent boundary layers. {\em
  Phil. Trans. R. Soc. A\/} {\bf 365}, 755--770.

\bibitem[Nagib {\em et~al.\/}(2024)Nagib, Vinuesa \& Hoyas]{nagib2024utilizing}
{\sc Nagib, H., Vinuesa, R. \& Hoyas, S.} 2024 Utilizing indicator functions
  with computational data to confirm nature of overlap in normal turbulent
  stresses: logarithmic or quarter-power. {\em Phy. Fluids\/} {\bf 36}, 075145.

\bibitem[Ono {\em et~al.\/}(2023)Ono, Furuichi \& Tsuji]{Ono2023}
{\sc Ono, M., Furuichi, N. \& Tsuji, Y.} 2023 Reynolds number dependence of
  turbulent kinetic energy and energy balance of 3-component turbulence
  intensity in a pipe flow. {\em Journal of Fluid Mechanics\/} {\bf 975}, A9.

\bibitem[Ono {\em et~al.\/}(2022)Ono, Furuichi, Wada, Kurihara \&
  Tsuji]{Ono2022}
{\sc Ono, M., Furuichi, N., Wada, Y., Kurihara, N. \& Tsuji, Y.} 2022 Reynolds
  number dependence of inner peak turbulence intensity in pipe flow. {\em Phys.
  Fluids\/} {\bf 34}, 045103.

\bibitem[Panton {\em et~al.\/}(2017)Panton, Lee \& Moser]{Panton2017}
{\sc Panton, R.L., Lee, M. \& Moser, R.D.} 2017 {C}orrelation of pressure
  fluctuations in turbulent wall layers. {\em Phys. Rev. Fluids\/} {\bf 2},
  094604.

\bibitem[Perry {\em et~al.\/}(1986)Perry, Henbest \& Chong]{Perry1986}
{\sc Perry, A.E., Henbest, S. \& Chong, M.S.} 1986 A theoretical and
  experimental study of wall turbulence. {\em J. Fluid Mech.\/} {\bf 165},
  163--199.

\bibitem[Pirozzoli(2024)]{Pirozzoli2024}
{\sc Pirozzoli, Sergio} 2024 On the streamwise velocity variance in the
  near-wall region of turbulent flows. {\em Journal of Fluid Mechanics\/} {\bf
  989}, A5.

\bibitem[Samie {\em et~al.\/}(2018)Samie, Marusic, Hutchins, Fu, Fan, Hultmark
  \& Smits]{Samie2018JFM}
{\sc Samie, M., Marusic, I., Hutchins, N., Fu, M.K., Fan, Y., Hultmark, M. \&
  Smits, A.J.} 2018 Fully resolved measurements of turbulent boundary layer
  flows up to {R}e$_\tau$ = 20000. {\em J. Fluid Mech.\/} {\bf 851}, 391--415.

\bibitem[Schlatter {\em et~al.\/}(2014)Schlatter, Li, Örlü, Hussain \&
  Henningson]{Schlatter2014}
{\sc Schlatter, P., Li, Q., Örlü, R., Hussain, F. \& Henningson, D.S.} 2014
  On the near-wall vortical structures at moderate {R}eynolds numbers. {\em
  Euro. J. Mech. B/ Fluids\/} {\bf 48}, 75--93.

\bibitem[Schlatter \& \"{O}rl\"{u}(2010)]{SO2010}
{\sc Schlatter, P. \& \"{O}rl\"{u}, R.} 2010 Assessment of direct numerical
  simulation data of turbulent boundary layers. {\em J. Fluid Mech.\/} {\bf
  659}, 116--126.

\bibitem[Smits(2022)]{Smits2022prize}
{\sc Smits, A.J.} 2022 Batchelor {P}rize {L}ecture: {M}easurements in
  wall-bounded turbulence. {\em J. Fluid Mech.\/} {\bf 940}, A1.

\bibitem[Smits {\em et~al.\/}(2011)Smits, McKeon \& Marusic]{Smits2011ARFM}
{\sc Smits, A.J., McKeon, B.J. \& Marusic, I.} 2011 High-{R}eynolds number wall
  turbulence. {\em Annu. Rev. Fluid Mech.\/} {\bf 43}, 353--375.

\bibitem[Sreenivasan(2018)]{Sreenivasan_mixing}
{\sc Sreenivasan, Katepalli~R.} 2018 Turbulent mixing: A perspective. {\em
  Proceedings of the National Academy of Sciences\/} {\bf 116}, 18175 -- 18183.

\bibitem[Sreenivasan \& Antonia(1997)]{SreenivasanAntonia}
{\sc Sreenivasan, Katepalli~R. \& Antonia, Robert~A.} 1997 The phenomenology of
  small-scale turbulence. {\em Annual Review of Fluid Mechanics\/} {\bf
  29}~(1), 435--472.

\bibitem[Townsend(1976)]{Townsend1976}
{\sc Townsend, A.A.} 1976 {\em The structure of turbulent shear flow\/}.
  Cambridge University Press.

\bibitem[Vallikivi {\em et~al.\/}(2015)Vallikivi, Ganapathisubramani \&
  Smits]{vallikivi2015}
{\sc Vallikivi, M., Ganapathisubramani, B. \& Smits, A.J.} 2015 Spectral
  scaling in boundary layers and pipes at very high {R}eynolds numbers. {\em J.
  Fluid Mech.\/} {\bf 771}, 303--326.

\bibitem[Vincenti {\em et~al.\/}(2013)Vincenti, Klewicki, Morrill-Winter, White
  \& Wosnik]{vincenti2013}
{\sc Vincenti, P., Klewicki, J., Morrill-Winter, C., White, C.M. \& Wosnik, M.}
  2013 Streamwise velocity statistics in turbulent boundary layers that
  spatially develop to high {R}eynolds number. {\em Exp. Fluids\/} {\bf 54},
  1629.

\bibitem[Willert {\em et~al.\/}(2017)Willert, Soria, Stanislas, Klinner, Amili,
  Eisfelder, Cuvier, Bellani, Fiorini \& Talamelli]{LongPipe2017}
{\sc Willert, C., Soria, J., Stanislas, M., Klinner, J., Amili, O., Eisfelder,
  M., Cuvier, C., Bellani, G., Fiorini, T. \& Talamelli, A.} 2017 Near-wall
  statistics of a turbulent pipe flow at shear {R}eynolds numbers up to 40 000.
  {\em J. Fluid Mech.\/} {\bf 826}, R5.

\bibitem[Wu \& Moin(2009)]{WuxiaohuaTBL2009}
{\sc Wu, X.H. \& Moin, P.} 2009 Direct numerical simulation of turbulence in a
  nominally zero-presure-gradient flat-plate boundary layer. {\em J. Fluid
  Mech.\/} {\bf 630}, 5--41.

\bibitem[Yamamoto \& Tsuji(2018)]{Yama2018}
{\sc Yamamoto, Y. \& Tsuji, Y.} 2018 Numerical evidence of logarithmic regions
  in channel flow at $\mathrm{R}{\mathrm{e}}_{\ensuremath{\tau}}=8000$. {\em
  Phys. Rev. Fluids\/} {\bf 3}, 012602.

\bibitem[Yao {\em et~al.\/}(2023)Yao, Rezaeiravesh, Schlatter \&
  Hussain]{Yao2023JFM}
{\sc Yao, J., Rezaeiravesh, S., Schlatter, P. \& Hussain, F.} 2023 Direct
  numerical simulations of turbulent pipe flow up to {R}$e_\tau \approx 5200$.
  {\em J. Fluid Mech.\/} {\bf 956}, A18.

\end{thebibliography}
\bibliographystyle{jfm}

\end{document}